\documentclass[10pt]{article}

\usepackage{amsmath}
\usepackage{amssymb}

\usepackage{graphicx}

\usepackage{cite}

\usepackage{color} 

\usepackage[labelfont=bf,labelsep=period,justification=raggedright]{caption}

\makeatletter
\renewcommand{\@biblabel}[1]{\quad#1.}
\makeatother

\date{}

\usepackage{amssymb}
\usepackage{graphicx}

\usepackage{url}

\usepackage{graphicx,url}
\usepackage{epsfig}

\DeclareGraphicsExtensions{.pdf,.png,.jpg,.eps}

\usepackage[tight,footnotesize]{subfigure}
\usepackage{verbatim}
\usepackage[latin1]{inputenc}  

\usepackage{algorithm}
\usepackage{algorithmic}

\usepackage{amsthm}

\usepackage{color}

\usepackage{paralist}

\begin{document}

\begin{flushleft}
{\Large
\textbf{DANCE: A Framework for the Distributed Assessment of Network Centralities}
}
\\
Klaus~Wehmuth,$^{\ast}$ 
Ant\^{o}nio Tadeu~A.~Gomes, 
Artur~Ziviani
\\
National Laboratory for Scientific Computing (LNCC), Petr\'{o}polis, RJ, Brazil
\\
\end{flushleft}

\section*{Abstract}
The analysis of large-scale complex networks is a major challenge in the Big Data domain. 
Given the large-scale of the complex networks researchers commonly deal with nowadays, the use of localized information~(i.e. restricted to a limited neighborhood around each node of the network) for centrality-based analysis is gaining momentum in the recent literature. In this context, we propose a framework for the Distributed Assessment of Network Centralities~(DANCE) in complex networks. DANCE offers a single environment that allows the use of different localized centrality proposals, which can be tailored to specific applications. 
This environment can be thus useful given the vast potential applicability of centrality-based analysis on large-scale complex networks found in different areas, such as
Biology, Physics, Sociology, or Computer Science.
Since the localized centrality proposals DANCE implements employ only localized information, DANCE can easily benefit from parallel processing environments and run on different computing architectures. To illustrate this, we present a parallel implementation of DANCE and show how it can be applied to the analysis of large-scale complex networks using different kinds of network centralities. This implementation is made available to complex network researchers and practitioners interested in using it through a scientific web portal.


\section{Introduction}
\label{sec:intro}

The analysis of large-scale complex networks is a major challenge in the Big Data domain. 
In this context, network centrality is a key concept in the analysis of complex networks.
Roughly speaking, the concept of network centrality deals with the assessment of the relative importance
of nodes in a given network following some criteria. 
This concept has been around for a long time and different ways of measuring network centrality
have been proposed for decades~\cite{Sabidussi1966},
each of them suited to assess centrality from a different point of view, targeting a different goal.
Nevertheless, many of the traditional network centrality definitions 
have a high computational cost and require full knowledge of the network topology to be computed.
Later we provide a brief background on traditional network centralities discussing
their computational costs.

The high computational cost and the requirement of full knowledge of the network topology may prevent 
the application of centrality-based analysis to large-scale complex networks. 
This is a key challenge in Big Data analysis given the prevalence and scale of the complex networks we face today.
Just to keep in the realm of techno-social networks~\cite{Vespignani2009}, the Internet routing structure, online social networks, P2P networks, Internet of Things~(IoT), and content distribution networks represent some examples of networks that impose important challenges for the direct application of costly traditional centralities for network analysis because of their large-scale. Similar examples exist in Biology, Physics, Sociology, and in several other fields that deal with large-scale complex networks.

Inline with this trend, recent research in network science has been dedicated to dealing with centralities in large-scale
complex networks. For instance, some recent efforts aim at optimizing the way traditional centralities are
calculated~\cite{Shi2011,Baglioni2012,Ishakian2012,Lee2012}. These methods are, however, still computationally expensive, in particular for centralities based on computing the shortest-paths between all node pairs.
Moreover, such methods still require full knowledge of the network topology to compute centrality, hindering their
applicability to large-scale networks where such information is unavailable and a distributed implementation is required.
Alternatively, some previous work proposes methods to distributively assess network centrality 
without requiring full knowledge of the network topology based on localized information~\cite{Lehmann2003,Everett2005,Nanda2008,Ercsey-Ravasz2010,Wehmuth2012b,Pant13:Local,Wehmuth-comnet}.
In particular, Marsden~\cite{Marsden2002} shows empirical evidence 
that localized centrality measures computed for one-hop neighborhoods are highly correlated to a global centrality measure.
In general, however, these proposals for the distributed evaluation of network centralities disregard issues concerning the optimization
of their implementations, therefore it is hard to compare their performances as well as to 
extend their applicability (e.g., by evaluating the impact of improving the locality they consider). These are the challenges we address in our work as we describe in the following.

In this paper, we propose DANCE~(\textit{Distributed Assessment of Network Centralities}),
a framework to distributively assess network centralities based only on localized information restricted to a given neighborhood around each node in the network.
DANCE allows extending in a practical and efficient way the notion introduced by Marsden~\cite{Marsden2002} 
of using localized centrality measures to assess global centrality measures
in large complex networks.
The DANCE framework offers a single environment that allows the use of different localized centrality proposals, which can be tailored to a specific application. 
To that end, DANCE can be customized for calculating different centralities by only changing a single function that receives a limited neighborhood
around each node and returns a real number, thus allowing node ranking. The definition of this function determines which network centrality is assessed through the use of the proposed framework. 
DANCE is specialized in the distributed assessment of network centralities based on localized information, thus making it simpler and more straightforward to use as compared with 
other general-purpose graph processing environments, such as GraphLab~\cite{graphlab}, Pregel~\cite{pregel}, or Giraph.\footnote{\url{http://giraph.apache.org/}} 
This outcome is quite useful given the vast potential applicability of centrality-based analysis to large-scale complex networks present in different areas of knowledge.
 
The DANCE contribution is threefold.
First, to the best of our knowledge, the DANCE framework is the first practical environment specifically focused on analyzing, extending, and investigating localized centrality evaluations to approximate global centrality-based network properties. DANCE provides a single environment where it is easy to analyze and extend existing proposals for localized centrality assessment, thus also enhancing the comparability between such proposals. Such a framework also allows easier 
implementation of new proposals for the localized centrality assessment that approximate some global network property.
Second, previous individual proposals for the distributed evaluation of network centrality disregard issues concerning their optimization for parallelized computational environments.
In contrast, given that the localized centrality proposals DANCE implements employ only localized information, DANCE can easily benefit from parallel processing environments and run on different computing architectures. 
In this way, DANCE considers different parallel strategies for an improved implementation within parallel environments.
Third, an implementation of the DANCE framework offering localized centrality assessments that approximate some well-known global centrality measures is made available to complex network researchers and practitioners interested in using it through a scientific web portal~(\url{http://www.lncc.br/sinapad/DANCE})~\cite{cpe-sinapad}.


This paper proceeds as follows. 
Section~\ref{sec:bkg} briefly reviews the definitions of the most traditional network centralities.
Section~\ref{sec:found} introduces DANCE and discusses its properties.
In Section~\ref{sec:implement}, we present the reference architecture of DANCE and discuss implementation issues to run DANCE
in parallel scenarios. Performance evaluation is presented in Section~\ref{sec:res}. 
Finally, Section~\ref{sec:conc} concludes the paper and discusses future work.

\section{Background on network centralities}
\label{sec:bkg}

In this section, considering a network with $n$ nodes and $m$ edges,
we briefly review the definitions of the most traditional centralities
used in network analysis:

\begin{itemize}

\item \emph{Degree centrality} considers the degree of a node as the centrality value of that node.
Calculating the degree centrality for all the nodes in a network
takes $O(n^2)$ in a dense adjacency matrix representation of the network, which becomes $O(m)$ in a sparse graph.

\item \emph{Betweenness centrality}
is based on the idea that a node is important when many shortest paths pass through it, making that node
an intermediary between many nodes.
It is thus defined at each node as the fraction of shortest paths between all node pairs that pass through each node.
Calculating betweenness centrality involves computing the shortest paths between all pairs of nodes in the networks and, 
in the general case, this requires $O(n^3)$ time.
For unweighted graphs, Brandes algorithm~\cite{Brandes2001}
is the current state-of-the-art and takes $O(nm)$ time for an exact computing of betweenness centrality for all nodes. 


\item \emph{Closeness centrality}
is based on the idea that a node is important if it is close to all other nodes in the network.
The closeness centrality of a node is computed taking the inverse of the sum of the shortest distances in hops to all other nodes.
Calculating closeness centrality also requires computing the shortest paths between all pairs of nodes in the network.
Hence, the same algorithms apply with the same time complexities as in the case for betweenness centrality.

\item \emph{Eigenvector centrality}
is determined by the eigenvector associated to the largest eigenvalue of the network adjacency matrix.
The values of the eigenvector entries define the centrality value of each node. The basic rationale behind eigenvector
centrality is that connections from nodes with higher degree contribute more to the centrality of each node than connections
from nodes with lower degree. 
In the general case, the traditional algorithm for this
takes $O(n^3)$ time, with possible improvements in performance in particular cases. 
For instance, Google's PageRank is a variant of this network centrality.

%

\end{itemize}

Calculating these traditional network centralities is in general computationally expensive, particularly
in the case of those based on all shortest paths in the network, such as the betweenness and closeness centralities.
Although Brandes algorithm~\cite{Brandes2001} improves significantly the performance for unweighted graphs,
it is still quadratic in time for sparse networks where~$m=O(n)$. 
Given such complexity and the very large
complex networks we face nowadays in different areas, 
distributively evaluating network centralities based only on localized information becomes increasingly important.
This is the key motivation behind research on distributed solutions that can evaluate
network centralities based only on localized information. 


\section{The DANCE framework}
\label{sec:found}

In this section, we introduce the DANCE framework. First, we present some key definitions
behind DANCE. Second, we show how
DANCE is structured for the distributed assessment of network centralities. Finally, we present
the basic architectural components of the DANCE framework and we also provide some examples of localized computations that
can be applied to the limited neighborhood around each node to yield an assessment of a global property in network
analysis. 

\subsection{Key definitions}
\label{sub:bg}

We consider a network as equivalent to an \emph{undirected} finite simple graph $G=(V,E)$, where $V$ is the set of nodes and $E$ the set of edges. The distance between two nodes in the network is defined as the number of hops in the shortest path connecting these nodes.
The radius $r$ of a graph~$G$ is equivalent to the minimum eccentricity of any node, {\it i.e.}, $r =\min_{i \in V}(\max_{j \in V} d(i,j))$ where $d(i,j)$ is the shortest path distance between nodes~$i$ and~$j$. In the following, we introduce key definitions upon which DANCE is built: 

\begin{itemize}

\item Neighborhood -- We define the neighborhood of a node~$i$ with radius~$h$ as the set of nodes that have distance to node~$i$ less or equal than~$h$, and the 
set of edges that are incident to at least one of those nodes. This neighborhood is thus 
denoted as $H_h^i = (V_h^i, E_h^i)$, where~$i$ is the central node, $h$ is the radius of the neighborhood, 
$V_h^i$ is the set of nodes whose distance to~$i$ is less or equal to~$h$ and $E_h^i$ is the set of edges 
adjacent to at least one node of $V_h^i$. 
The notation $j \in H_h^i$ is used to represent $j \in V_h^i$, meaning that the node $j$ belongs to the neighborhood $H_h^i$. 

\item Set of all neighborhoods -- For every network $G=(V,E)$ and every $h\geq0$ there is a set $\mathbb{H}_h = \{H_h^i~|~i \in V\}$, containing 
all neighborhoods of radius $h$ in $G$.
It follows from this definition that the number of elements in  $\mathbb{H}_h$ is equal to the number of nodes in $G$, since every 
neighborhood $H_h^i$ is uniquely characterized by its central node~$i$. More formally, there is a function $f_n: V \to \mathbb{H}_h$ that associates every node $i$ in the network to its neighborhood $H_h^i$.

\item Classifier function -- A classifier function (or classifier for short)  is defined as a function 
$f_c: \mathbb{H}_h \to \mathbb{R}$ 
that takes neighborhoods into real numbers, creating an ordered partition upon the set of nodes~$V$ using only localized information from the neighborhood around each node.
In this way, the classifier function partitions the set of nodes~$V$ into \emph{equivalence classes}, hence the denomination classifier. 
The resulting set of equivalence classes, each associated with a real number, enables the {\it ranking} of the network nodes. 
This node ranking establishes the relative importance of the nodes---i.e., their centrality---in the network.
Therefore, for each specific classifier function adopted within DANCE, a different centrality notion may be obtained.


\end{itemize}

\subsection{Distributed assessment of network centrality}
\label{sub:cent_det}

DANCE works by taking a given integer radius $h$ and a classifier function $f_c$. 
First, DANCE determines the set $\mathbb{H}_h$ of all neighborhoods 
of radius $h$. Then, DANCE uses a selected classifier function $f_c$ to assess the centrality of each neighborhood and, in consequence, of each node in the 
network. Clearly, the choices of the radius $h$ of the neighborhoods as well as of the classifier 
function are crucial for the method to work properly in assessing how central nodes are for a given goal.


We thus remark that, for DANCE to work properly, the
radius~$h$ for determining the neighborhoods should balance different
aspects. On the one hand, $h$~should be large enough
so that the neighborhoods are representative as approximations of the network. On the other hand, $h$~should be small enough so that local topological properties 
can be translated into the neighborhoods, making it possible to discriminate them in order to create the basis for a centrality ranking. 
Although the choice of~$h$ is application-specific, the literature~(e.g., in~\cite{Everett2005,Wehmuth-comnet}) shows that small values, typically $h=1$ or $h=2$, yield good results in distributively assessing network centralities for different kinds of complex networks.

In addition to a suitable choice of the radius $h$ of the neighborhoods, it is also key to choose a suitable classifier function~$f_c$
for the global property to be approximated. This choice is, of course, also application-driven. 
Clearly, a poor choice for the classifier can lead to an inappropriate result. For 
instance, if the zero function ($f_z(v) = 0$ for all $v$) is chosen as a classifier, all neighborhoods are evaluated as having value zero, 
making all nodes equivalent regardless of the radius used to create the neighborhoods.
Therefore, classifiers should be functions that use the localized characteristics present in the neighborhoods to discriminate 
them creating a ranking that reflects the relative importance of the nodes in the network. 
In Section~\ref{sub:impl}, we present some practical examples of such classifier functions.

\subsection{DANCE framework usage and classifiers}
\label{sub:impl}

\begin{figure}[!h]
\centering
 \includegraphics[width=0.75\columnwidth,keepaspectratio=true]{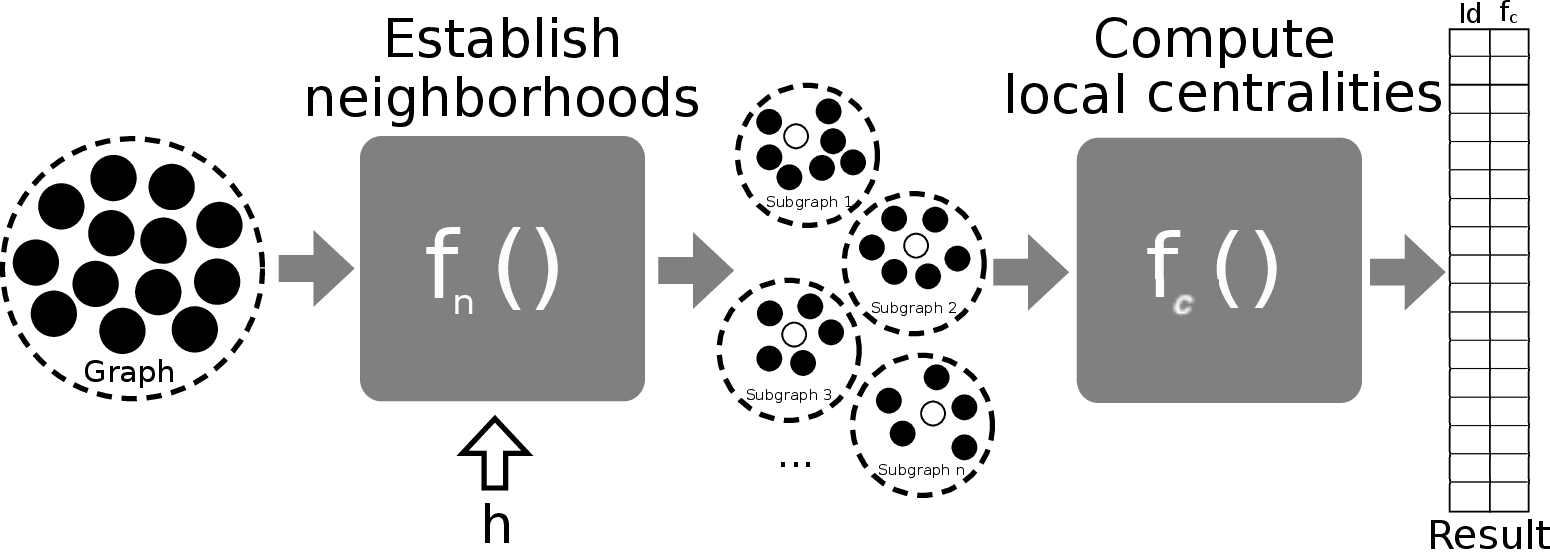}
 \caption{The DANCE framework.}
  \label{fig:Fram_DANCE}
\end{figure}

Considering the previous discussion, the DANCE framework consists of two main components: (i)~the determination of 
the neighborhood with radius~$h$ of each node~$i$ of the network~(using the previously defined $f_n$ function); and (ii)~the determination of the centrality value~$v_i = f_c(H_h^i)$ of each 
node~$i$ using the classifier function~$f_c$. 
Figure~\ref{fig:Fram_DANCE} presents how these components 
are integrated within the DANCE framework.


The function $f_n$ is applied to each node in the network, 
rendering for each such node a subgraph containing the node's neighborhood of radius~$h$.
%
%
The determined classifier function $f_c$ is then applied to each of such subgraphs, 
in order to get the localized centrality value of the node corresponding to each of such subgraphs and establish its centrality ranking. 
The classifier is the key component where the DANCE framework can be tailored to different notions of network centrality. 
Note that different classifier functions are available at the DANCE framework for the user to choose and these
may be easily changed by already implemented ones or new ones for evaluation. 
In the following, we present a few examples of classifier functions and briefly discuss their expected results. (The particular characteristics
and performance evaluation of each considered localized centrality are discussed in the respective paper which has proposed it.)


\begin{itemize}

\item \emph{Ego-betweenness}. A classifier that locally computes the traditional betweenness centrality of each local neighborhood replicates the 
ego-betweenness centrality proposed by Everett and Borgatti in~\cite{Everett2005}, when used with radius $h=1$, which is shown to be highly correlated to the traditional betweenness centrality.
When used with radius $h > 1$,  it would extend the 
concept of ego-betweenness to larger neighborhoods that can better represent larger complex networks, making it possible to use this localized centrality notion 
in large-scale networks.

\item \emph{Localized bridging centrality (LBC)}. Another possible classifier is to define $f_c(i) = BC(i) \times \beta(i)$, where $BC(i)$ is the traditional betweenness 
centrality of the central node $i$ calculated on its neighborhood and $\beta(i)$ is a bridging coefficient~\cite{Hwang2008} defined as 


\begin{equation}
\beta(i) = \frac{\sum_{j \in H_1^i}d_j}{d_i},
\end{equation}

\noindent
where $d_i$ is the degree of node~$i$.
This classifier, when used with radius $h=1$, replicates the Localized Bridging Centrality~(LBC) proposed by Nanda and Kotz~\cite{Nanda2008}.
When used with radius $h > 1$, it would again extend this centrality concept making it possible to use it in a wider range of networks than those 
originally proposed. The goal here is to find bridges in the network, i.e. nodes that connect two densely connected components in the network.

\item \emph{Volume-based}. In a previous work~\cite{Wehmuth-comnet},
we propose a classifier named DACCER that calculates the volume of each neighborhood, i.e. the sum of the degrees of all nodes belonging to the considered neighborhood. From the definitions 
of neighborhood and volume, it can be seen that with radius $h=0$ this classifier generates the traditional degree centrality. However, 
with $h > 0$, the centrality ranking generated by this classifier correlates strongly with the ranking generated by the traditional closeness 
centrality, as shown in~\cite{Wehmuth-comnet}.
Kim and Yoneki~\cite{Kim2012a} recently
proposed an extension to the volume-based classifier proposed in~\cite{Wehmuth-comnet} that better approximates the closeness centrality values instead of the node ranking.



\end{itemize}

\section{DANCE implementation}
\label{sec:implement}

\begin{figure}[!h]
\centering
 \includegraphics[width=0.75\columnwidth,keepaspectratio=true]{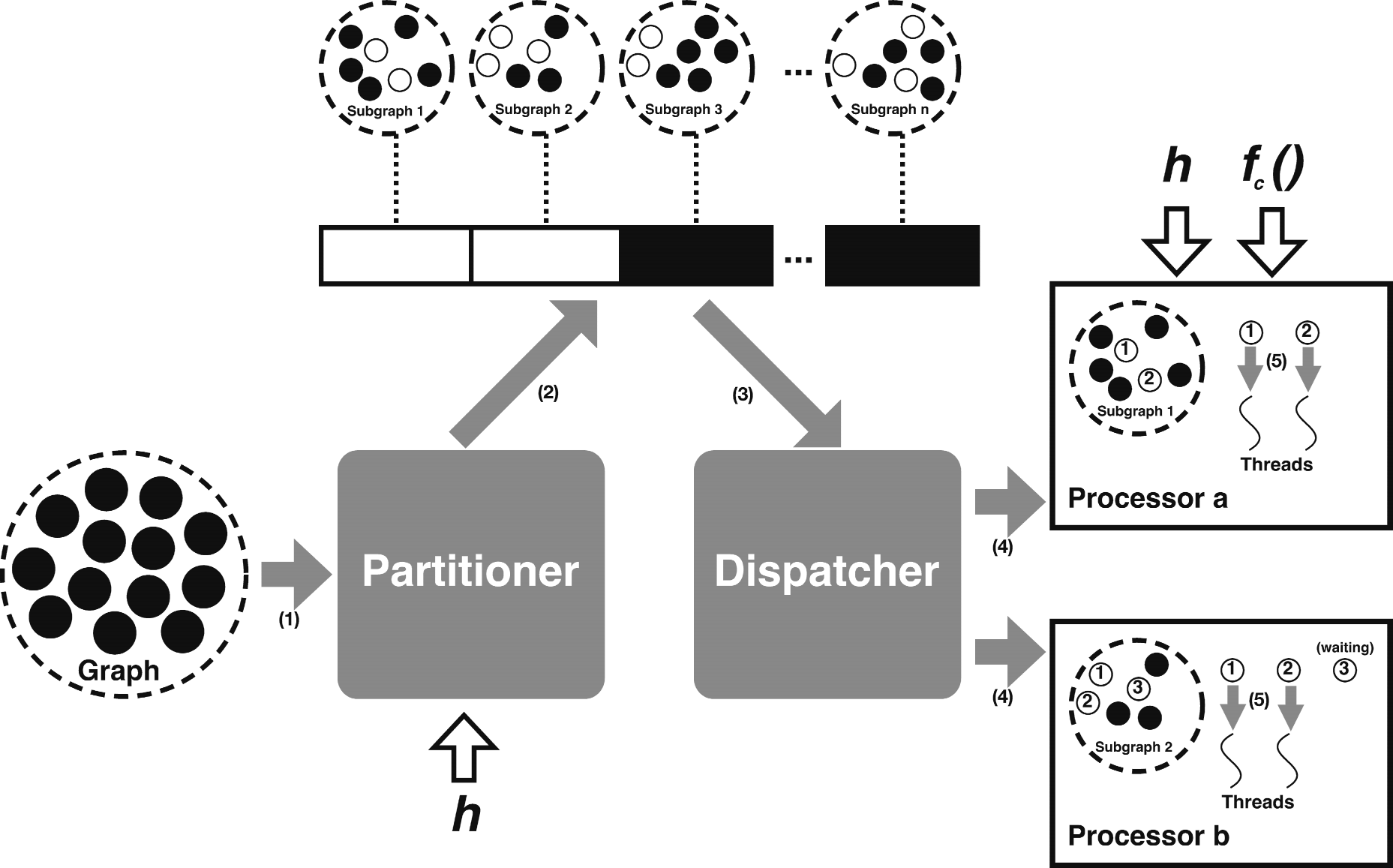}
 \caption{Reference architecture for the parallel implementations of DANCE.}
  \label{fig:Arch_DANCE}
\end{figure}

Figure~\ref{fig:Arch_DANCE} shows a reference architecture for the parallel implementations of DANCE.
The two key components of this architecture are the \emph{partitioner} and the \emph{dispatcher}:

\begin{itemize}

\item The partitioner is responsible for generating a set of $n$ subgraphs from the original network graph (steps (1) and (2) in Figure~\ref{fig:Arch_DANCE}).
These subgraphs are stored in a \emph{subgraph queue} to be subsequently retrieved and processed in parallel by a set of processors in the underlying computing infrastructure. 
Importantly, these subgraphs may be also generated in parallel. 
For each subgraph $i$ in the subgraph queue, $1 \leq i \leq n$, the partitioner also elects a subset $V_i$ of nodes to compute the centrality from (the so-called \emph{origin} nodes---represented in white in Figure~\ref{fig:Arch_DANCE}).

\item The dispatcher is responsible for retrieving the subgraphs from the subgraph queue and submitting them for being processed in parallel in the underlying computing infrastructure (steps (3) and (4) in Figure~\ref{fig:Arch_DANCE}).
Once within a processor in the infrastructure, a subgraph is processed as follows: 
\begin{inparaenum}[(i)]
\item the origin nodes are stored in a \emph{node queue}, which is implemented as a priority queue in descending order by the node degree;
\item a pool of threads within the processor (one thread per core) keeps on dequeuing origin nodes from the priority queue (step (5) in Figure~\ref{fig:Arch_DANCE}) and computing their desired centrality according to the provided classifier function $f_{c}$.    
\end{inparaenum}

\end{itemize}


Note that the partitioner and dispatcher components presented in Figure~\ref{fig:Arch_DANCE} are not related to the functions~$f_c$ and~$f_n$, defined in Section~\ref{sub:bg} and shown in Figure~\ref{fig:Fram_DANCE} as part of the conceptual view of the DANCE framework.
To date, we have implemented and deployed the reference architecture in Figure~\ref{fig:Arch_DANCE} in two different parallel computing infrastructures: 
\begin{inparaenum}[(i)]
\item in a shared memory machine based on NUMA (Non-Uniform Memory Access) architecture by employing the Open Multiprocessing~(OpenMP) API;\footnote{\url{http://openmp.org}} and
\item in a computing cluster by using the Message Passing Interface (MPI).\footnote{\url{http://www.mpi-forum.org}} 
\end{inparaenum}

In the first implementation~(OpenMP), there is no partitioner and dispatcher. 
The whole graph $G$ is loaded to the machine's memory at once, and a pool of OpenMP threads is continuously served by the node queue until no origin nodes remain to be processed by these threads.
This is the simplest and most efficient implementation, but NUMA-based hardware is quite expensive and usually not so promptly available.
This implementation can be also used in hardware based on symmetric multiprocessing~(SMP), but the considerably smaller amount of memory available in this type of hardware hinders the use of this implementation for very large graphs. 

In the second implementation~(MPI), the partitioner and dispatcher are implemented in a single, master MPI process with process number $\#0$.\footnote{Here we do not employ the more common term ``MPI rank'' to refer to such number to avoid confusion with the concept ``node rank'' used in complex network analysis.}
Each processor $1 \leq i \leq p$ in the cluster runs an MPI process with process number $\#i$ that receives from MPI process $\#0$ a specific subgraph $i$ and a subset $V_i$ of origin nodes.
Each MPI process $\#i$ then implements a pool of OpenMP threads (like in the first implementation) to calculate the desired centrality of its subset of origin nodes in parallel. 
In this implementation, $n=p$, i.e.\ there is no subgraph queue.
An MPI process $\#x$ may communicate with an MPI process $\#y$ whenever $\#x$ needs neighborhood information about a node stored in $\#y$ to compute the desired centrality.
The strategies for partitioning the graph among the MPI processes may lie in-between two extreme cases:
\begin{itemize}
\item \textbf{No node replication.} The subset $V_i$ of origin nodes equals the set of vertices in the subgraph stored by a processor;
\item \textbf{Full node replication.} The set of vertices in the subgraph equals the set of vertices in the original graph. 
\end{itemize}
The `no node replication' strategy requires the smallest amount of memory from the processors, but leads to the largest communication overhead between the MPI processes.
The `full node replication' strategy is equivalent to a completely `shared-nothing' implementation, requiring no communication between the MPI processes but demanding a larger amount of local memory from the processors.
To date, we have implemented both extreme strategies. 
We have recently progressed work on a partitioning strategy that allows memory usage to be traded for some acceptable increase in the total execution makespan.
The main idea behind this partitioning scheme relies on a computationally-cheap partitioning phase that partially replicates the graph throughout the MPI processes in a way that interprocess communication during the  centrality computation phase is minimized or not necessary at all.

All the aforementioned implementations are made available to complex network researchers and practitioners interested in using the DANCE framework through
a scientific web portal\footnote{\url{http://www.lncc.br/sinapad/DANCE}}~\cite{cpe-sinapad}.
Such portal offers implementations for all classifier functions described in this paper.

%

%

\section{Performance evaluation}
\label{sec:res}
As already mentioned, the particular efficiency of each localized classifier in assessing a global property of the network is shown in the papers that proposed each classifier~\cite{Everett2005,Hwang2008,Wehmuth-comnet}.
In contrast, the main evaluation objective in this work is to analyze how the proposed framework behaves in processing network centralities. To achieve this goal, we present experiments conducted on a single shared memory machine with 8 CPU cores (an Intel Xeon Dual Quad Core 2.27GHz with 48GB RAM).
Our experiments aim at showing that fairly large networks can be processed even in a modest computational environment and that the overall construction of the framework leads to a runtime structure that provides a good load balance, optimizing the use of the resources allocated to the centrality assessment workload.

For evaluating the performance of the parallel implementations of DANCE, we compare a pure OpenMP to a pure MPI 
implementation. 
We consider both a naive method for allocating nodes to be processed by each CPU core and a 
more elaborated one which is meant to improve the overall load balance among the CPU cores.
In these evaluations, we use three traces of complex networks:
\begin{inparaenum}[(i)]
\item AS-Skitter network~\cite{skitternet} with 1,696,415~nodes and 11,095,298~edges;
\item Youtube network~\cite{youtubenet} with 1,134,890 nodes and 2,987,624~edges;
\item Actors network~\cite{Barabasi1999} with 374,511~nodes and 15,014,850~edges.
\end{inparaenum}


\subsection{Parallel strategies and performance}

\begin{figure}[!h]
 \centering
 \subfigure[DACCER]{
 \includegraphics[width=0.45\columnwidth,keepaspectratio=true]{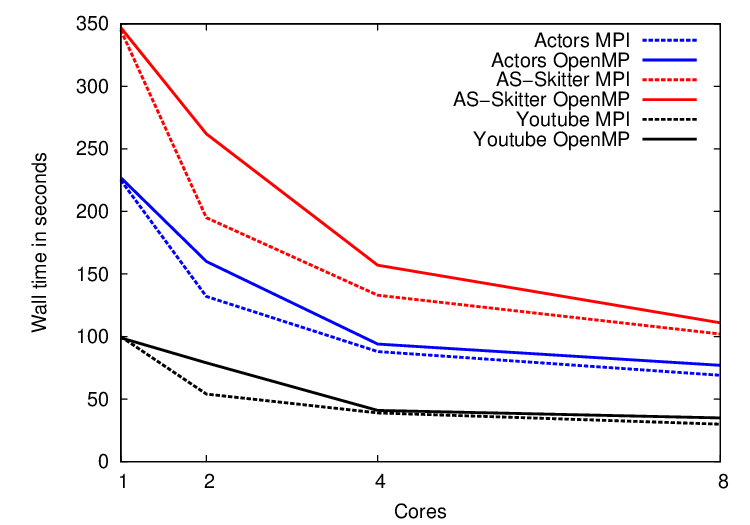}
 \label{fig:DAC2_Times}}
 \subfigure[Ego Betweenness]{
 \includegraphics[width=0.45\columnwidth,keepaspectratio=true]{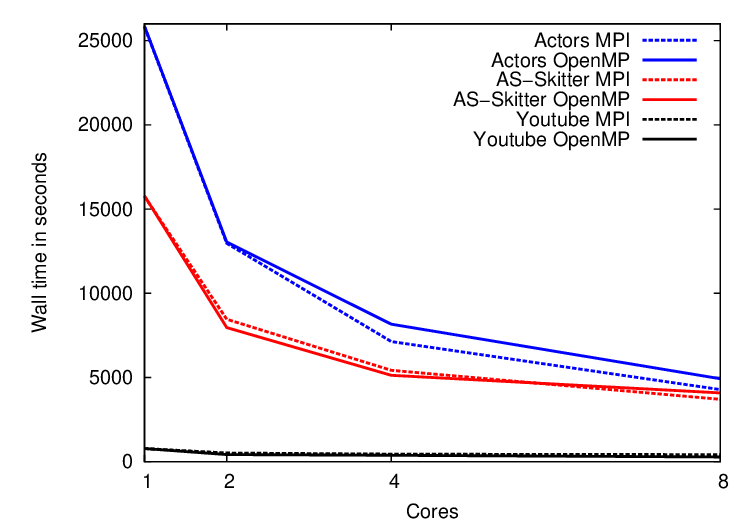}
 \label{fig:EBT1_Times}}
 \caption{Processing time comparison for MPI and OpenMP.}
 \label{fig:MPI_OMP}
\end{figure}

Figure~\ref{fig:MPI_OMP} shows the execution time comparison between the pure OpenMP and pure MPI implementation. It is important to remember that in this study both MPI and OpenMP were run on the same shared memory machine. The experiment compares the times needed to process the three considered networks using 1, 2, 4, and 8 cores. 
Figure~\ref{fig:DAC2_Times} shows the results obtained for the DACCER classifier~\cite{Wehmuth-comnet}, whereas Figure~\ref{fig:EBT1_Times} shows the results obtained for ego-betweenness classifier~\cite{Everett2005}. 
MPI outperforms OpenMP in this case.
This can be explained by the fact that there is no inter-process (or inter-thread) communication happening in this application, since each process (thread) can work completely independent of all others. This is a strong indication of good behavior of such implementation in really distributed memory environments.
Further, in the OpenMP implementation, a couple of critical regions have been used to assure thread-safety on the operations of getting nodes from the network to process, as well as writing the results obtained. Since this is not necessary in the MPI implementation, this justifies the observed advantage.
Moreover, the DACCER classifier is very lightweight when compared to the ego-betweenness classifier. (Observe the Y-axis scales in both figures.) Related with this observation, notice in Figure~\ref{fig:EBT1_Times} that the Actors network took longer to be processed than the network AS-Skitter, which is actually much larger. This happens because the Actors network is much denser than the AS-Skitter one, making the neighborhoods obtained on it much larger. Since, in the case of ego-betweenness, most of the cost is associated with running the classifier function, this causes the Actors network to be the most expensive one to evaluate. It should also be noticed that the times considered to build Figure~\ref{fig:MPI_OMP} are the wall time needed for running the whole application, including setup, reading the network from disk and writing the results back to disk. This makes clear that the DANCE framework is composed of an embarrassingly parallel part in which the centralities are calculated as well as other setup and result sections that are not easily parallelizable. This accounts for the diminishing speed-up as the number of cores increases and also explains why on Figure~\ref{fig:EBT1_Times} the time needed to process the Actors network takes longer than the time needed for AS-Skitter network. Since the Actors network is considerably smaller than the AS-Skitter, the time spent on the non parallel section for setup and result writing is smaller for the Actors network than for AS-Skitter. Thus, when the number of cores increases, the Actors network has a better speed-up.


\subsection{Load balancing}

\begin{figure}[!h]
 \centering
 \subfigure[DACCER]{
 \includegraphics[width=0.45\columnwidth,keepaspectratio=true]{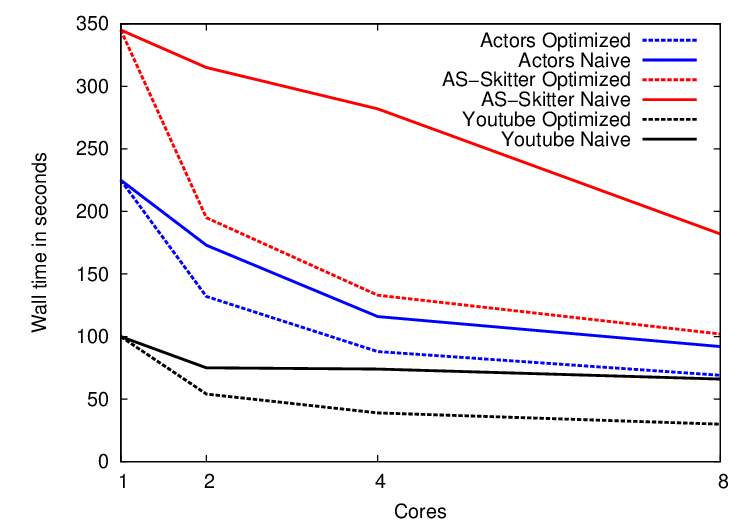}
 \label{fig:DAC2_Bal_Times}}
 \subfigure[Ego Betweenness]{
 \includegraphics[width=0.45\columnwidth,keepaspectratio=true]{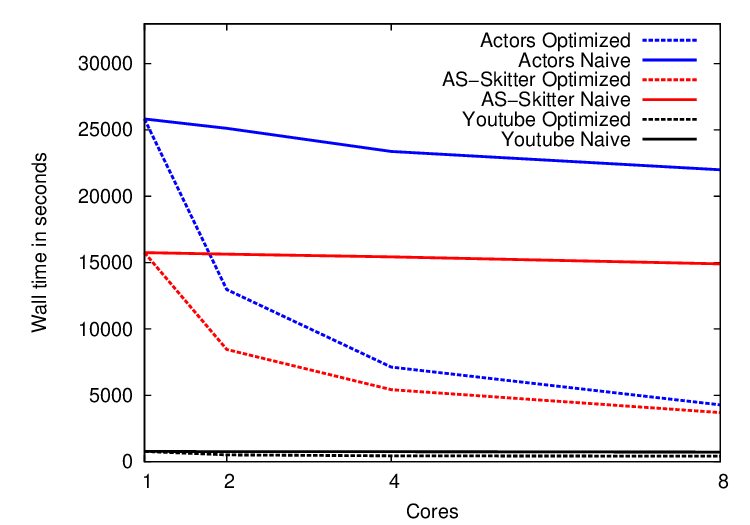}
 \label{fig:EBT1_Bal_Times}}
 \caption{Processing time comparison for naive and optimized load balancing.}
 \label{fig:Bal_Times}
\end{figure}

Next we compare the load balancing obtained by allocating the nodes to be processed by each node on a naive way and on an improved node allocation method. On the naive approach we just allocate a contiguous range of nodes to each processor core. This usually leads to a bad load balancing on the processing, because it is rather common that nodes with high degree end up grouped in a same region.  This is due to the sampling of network traces and the generation algorithm on synthetic networks, which tend to group nodes with high degree. Further, these high degree nodes tend to be connected to each other, causing a certain segment of nodes to have high volume neighborhoods. When such a segment is assigned to the same processor core, this core will have a much larger load than all others, leading to a poor load balancing. The improved node allocation method consists in sorting the nodes by degree and then allocating one by one in sequence to each core. This spreads the high degree nodes among the processing cores, leading to a better load balancing. This is a known strategy for improving load balancing, and as can be seen in 
Figure~\ref{fig:Bal_Times},
has good results when applied to DANCE.





\section{Conclusion}
\label{sec:conc}

In this paper, we present DANCE, a framework that allows the assessment of node centralities in a distributed way for large scale complex networks. 
DANCE can be tailored to compute a localized centrality at each node considering only a limited neighborhood around each node. 
This localized centrality value is calculated by a classifier function that receives a neighborhood and returns a real number, thus allowing node ranking. This makes it easy to extend the framework to calculate new centralities, which is achieved by simply implementing new classifier functions. 
Most complex networks of interest present \textit{small world} properties (\textit{i.e.}, small radius compared to network size), making DANCE applicable to them using small values for the radius \textit{h} of the local neighborhoods, making it practical for the analysis of these networks. 
Since DANCE implements localized centrality proposals which are based only on local information, DANCE can be easily parallelized and run on different computing architectures.
In this way, DANCE has been continuously improved to run efficiently on different computational environments, including shared-memory, distributed-memory, and shared-nothing architectures.

The current DANCE implementation is built for unweighted networks. However, since the centrality calculation itself is performed by the customizable classifier function, it can be easily extended for weighted networks by implementing classifiers capable of coping with this kind of network.

%
%
%
%
%

\section*{Acknowledgments}
This work was partially supported by the Brazilian Funding Agencies
CAPES, CNPq, FINEP, and FAPERJ.



\newpage
\section*{Figure Legends}

\end{document}